\documentclass{svproc}
\usepackage{amsmath,amsfonts}
\usepackage{algorithmic}
\usepackage{array}
\usepackage[caption=false,font=normalsize,labelfont=sf,textfont=sf]{subfig}
\usepackage{textcomp}
\usepackage{microtype} 
\usepackage{stfloats}
\usepackage{collcell}
\usepackage{url}
\usepackage{placeins}
\usepackage{verbatim}
\usepackage{orcidlink}
\usepackage{graphicx}
\usepackage{nomencl}
\makenomenclature

\begin{document}
\mainmatter              
\title{Design of an Axial Flux Permanent Magnet Eddy Current Brake for Application on Light Weight Motor Vehicles}
\titlerunning{Design of an AFPM-ECBS for Application on LMVs}  
%
\author{Shankar Ramharack\inst{*}\orcidlink{0000-0003-3759-7333}}
\authorrunning{Shankar Ramharack} 
%
\tocauthor{Shankar Ramharack}
\institute{The University of the West Indies, St. Augustine, Trinidad and Tobago,
\email{shankar.ramharack@gmail.com}}

\maketitle              
\begin{abstract} 
Axial flux permanent magnet designs are compact and becoming a attractive design for electric vehicles as an auxiliary braking system. This work develops the design of an Axial Flux Permanent Magnet Eddy current brake for application on Light Weight Motor Vehicles as guided by industry regulations. The work makes a link between the common Finite Element Method approach used in the literature and the Analytical approach previously done. The design is conducted to meet the torque requirements per wheel. The average torque over the operating range closest to the design requirement was used as the solution.

    
\end{abstract}
\nomenclature{\(\mu_0\)}{The permeability of the air.}
\nomenclature{\(\sigma\)}{The electrical conductivity of the eddy-current conducting plate.}
\nomenclature{\(\alpha\)}{The pole-arc to pole-pitch ratio of the magnets}
\nomenclature{\(p\)}{ The number of pole pair}
\nomenclature{\(\mathbf{E}\)}{ Electric Field}
\nomenclature{\(\mathbf{B}\)}{ Magnetic field density}
\nomenclature{\(\mathbf{A}\)}{ Magnetic vector potential}
\nomenclature{\(\mathbf{J_e}\)}{ Current density in eddy-current conducting plate}
\nomenclature{\(\phi\)}{ Electric scalar potential}
\nomenclature{\(\mathbf{H_0}\)}{ Curl function space}
\nomenclature{\(H^{1}_0\)}{ Function space for which 1st-order derivatives exist}
\printnomenclature

\section{Introduction}
Electromechanical braking systems may be either of Magnetorheological fluid, permanent magnet, electromagnetic type. Furthermore, the geometry of these systems may be linear or axial. For most automobile applications axial systems are preferred for their easy integration into existing friction brake systems. Permanent magnet systems are simpler to construct than electromagnetic systems and magnetorheological fluid systems but they often require external control systems to vary placement for braking. Eddy current-based braking systems do not perform well at low speeds and thus are often used as an auxiliary to an existing friction-based braking system. Hence this work intends to design a Permanent Magnet(PM) eddy current brake for auxiliary braking at high revolutions per minute(RPM).

\subsection{Motivation}
Eddy-current couplings (ECCs) with permanent magnets (PMs) have the advantages of high efficiency, energy-saving, low maintenance cost, simple installation, and adaptation to harsh environments \cite{1201540},\cite{Xioquan2021}. The eddy current brake is an auxiliary braking system that can be mounted on the driveshaft or propeller shaft shown in Fig 1. Although it can be integrated into a friction brake as seen in Figure \ref{fig8}, the physics is complicated and cannot be done in the project period. 
\begin{figure}[!t]
\centering
\includegraphics[width=3.5in]{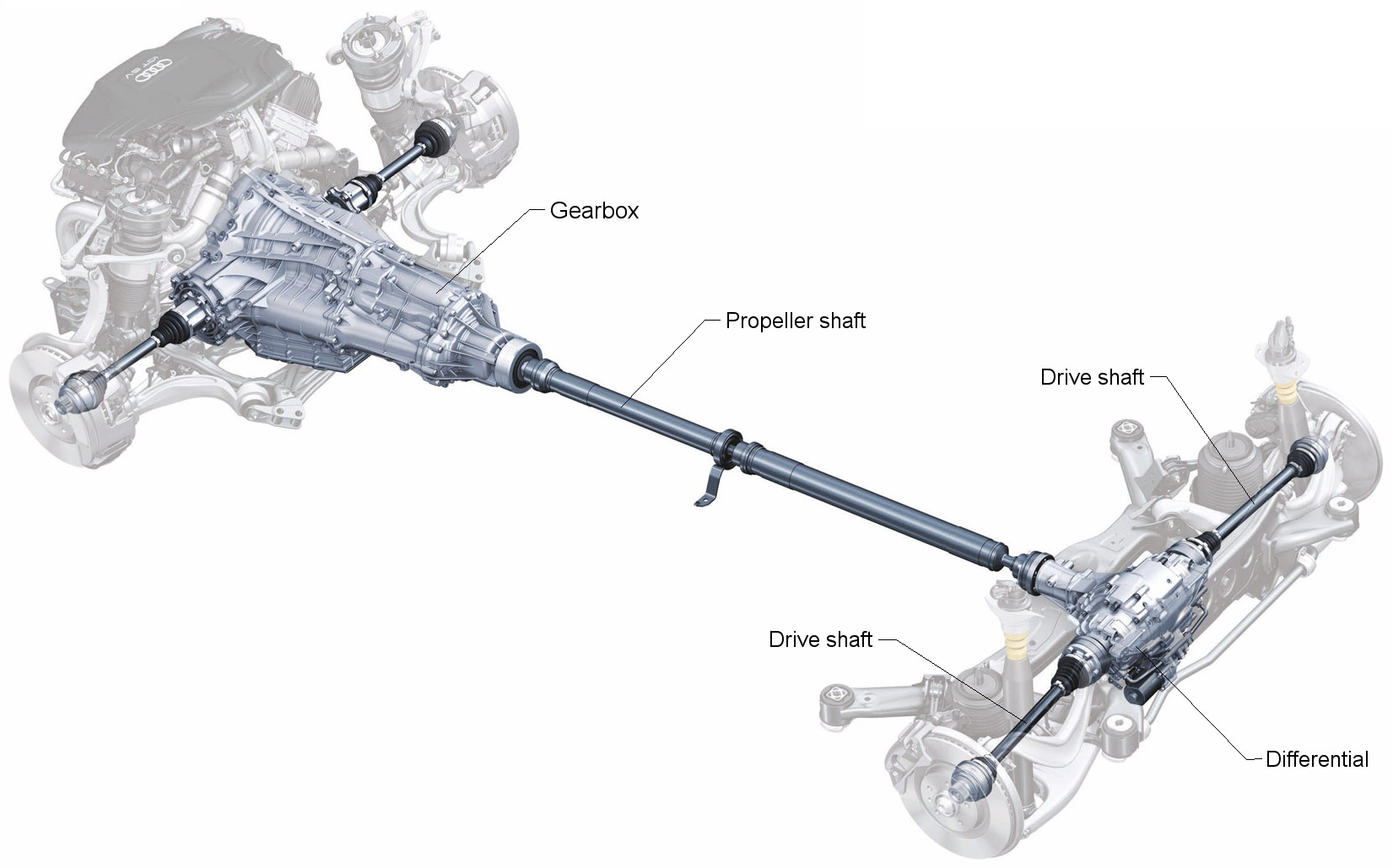}
\caption{Motion transfer mechanism of an automobile, Taken from \cite{takawane_2019_car}}
\label{fig1}
\end{figure}
\subsection{Background to the problem}
\begin{figure}[!t]
\centering
\includegraphics[width=3.5in]{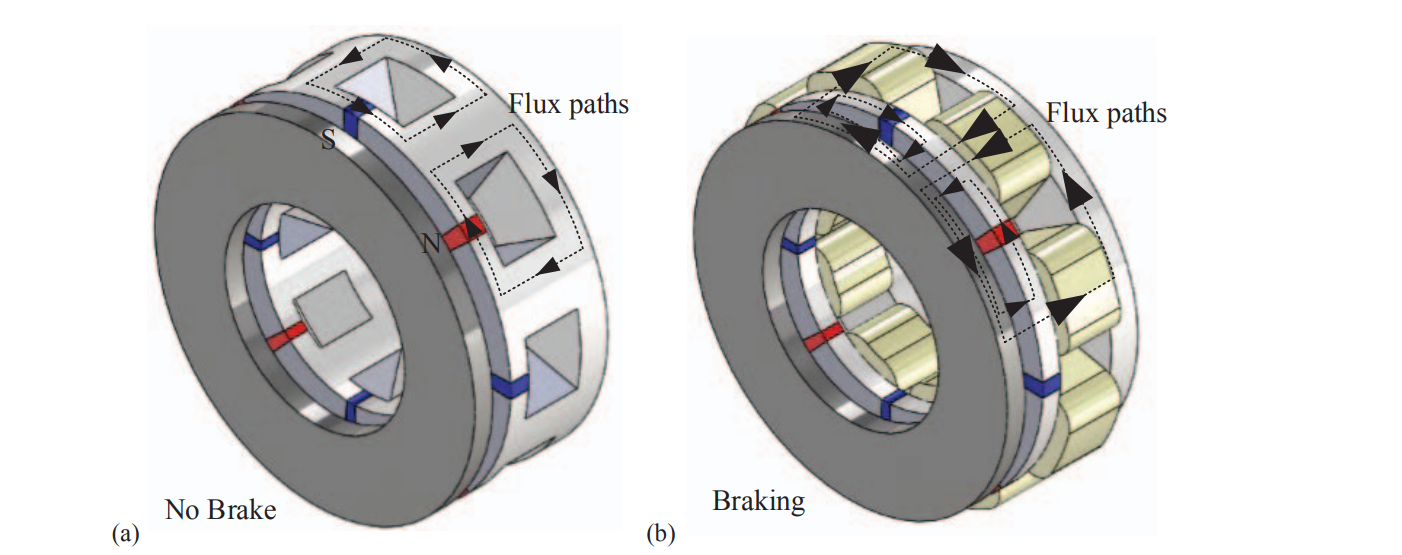}
\caption{Motion transfer mechanism of an automobile, Taken from \cite{takawane_2019_car}}
\label{fig2}
\end{figure}

Most Eddy Current Braking Systems(ECBS) \cite{putra_2021_mini,gulec_2016_design,shi_2013_study} utilize electromagnets to induce eddy currents in the rotor. Other systems use PM \cite{fontchastagner_2017_axialfield,fontchastagner_2018_design,shin_2013_analytical} to simplify construction but these systems then require complex control systems. An alternative is DC current assisted PM braking where an electromagnet is created by a dc current to short the flux paths as shown in Figure \ref{fig2}. 
Axial electromagnetic braking systems have been designed to reduce space and simplify construction \cite{waloyo_ubaidillah_tjahjana_nizam_koga_2019}. Furthermore, by using permanent magnets as the sole source of magnetic excitation, their is lower risk of electrical faults in the braking system. Electromagnetic braking systems in Light motor vehicles(LMVs) are expected to perform within specification reliably at high deceleration capacities. They are also expected to fit within existing braking housings to avoid major modification to the target vehicle. Designing Axial Flux Permanent Magnet braking systems have mostly been done using FEM. There have been attempts at analytical approaches to the design process but they do not consider typical design values in the parameter space such as those that are used in non-electric brakes. This is important as they make it easier to integrate into LMVs. Electric vehicles and Hybrids can greatly utilize electromagnetic braking systems to reduce maintenance costs,increase breaking capabilities and consequently, safety.

\subsection{Aims, Objectives and Scope}
This work aims to design an auxiliary braking system should provide braking Torque as required by a commercial braking regulations for 80\% of the average maximum speed of vehicles(per regulation). It should be able to meet extreme deceleration Torques and fit along the brake shaft or drive train.
\subsection{Outline of Paper}
Section I provides a background on the need for electromagnetic braking systems. Section II explains the operating principle of the axial flux permanent magnetic brake and develops a link between the Analytic design equations and the FEM verification model.  Section III covers the method used to traverse the parameter space of the design problem to arrive at a parameter combination that meet the design requirement. Section IV shows the results of the parameter space search as well as validation of the selected parameter's performance. Section V reviews the work done. 

\section{Background Theory}
The general theory of Eddy current braking is given in \cite{7394612}. In this work, the magnetization source is the permanent magnets. The conductive disc moves relative to the magnets inducing currents within it by Lenz's Law. These currents interact with the magnetic field to produce a Lorentz force retarding the rotation of the disc. 
\subsection{Simplifications and Assumptions}
To achieve the design requirements without expensive computational cost, the following simplifications and assumptions are made
\begin{enumerate}
    \item Thermal-magnetic effects are not considered in the paper
    \item It has been shown in \cite{delabarriere_2012_3d} that the curvature effects in the axial-flux actuators are generally a second-order phenomenon, and can be neglected without important errors. This assumption is critical to the Analytic approach used in the design.
    \item Steady state operation is assumed, I.E. the brake moves with constant slip speed
    \item The conducting plate has constant conductivity across it's geometry
    \item The induced currents in the back iron and the resulting back iron brake force contribution is neglected since it is negligible compared to the force on the main plate \cite{lubin_2015_3d}
    \item It is assumed that the required Torque required to brake the vehicle is distributed among the four wheels and any specialized force distribution can be done via magnetic shields and a control system which is out of the scope
\end{enumerate}

\subsection{Finite Element Equations}
The axial system studied in this paper is shown in Figure \ref{fig9}. In steady-state, the driving rotor rotates at speed $\dot{d}_1$ and the driven one at speed $\dot{d}_2$. Choosing the reference frame $(\mathrm{O},u_x,u_y,u_z)$ link to the first rotor, the first back iron is stationary, hence the the slip speed of the coupling is $\dot{d}  = \dot{d}_2 $ . All other components are stationary. The $z$ axis is defined as the axis of rotation and thus the rotation vector is $\dot{d} = \dot{d}u_z$. The domains and boundaries of the geometry are done using the same labels as the original authors\cite{fontchastagner_2017_axialfield}. The system can then be described using the Maxwell-Faraday equation in (1). For a full development of equation (1) see Appendix 2 and \cite{fontchastagner_2017_axialfield} for the original development.
\begin{equation}
\nabla \times (\bold{E} + \bold{v} \times \bold{B}) = \nabla \times (\bold{E} + (\bold{\dot{d}} \times \bold{p}) \times \bold{B} ) = 0
\end{equation}
The problem can then be accurately solved using the A-$\phi$ formulation as shown in \cite{fontchastagner_2017_axialfield}. By using a modified form of ohms Law, the eddy currents induced in the rotating disc can be determined using (2)
\begin{equation}
\bold{J_e} = \sigma(\ (-\nabla\cdot v) + (\bold{\dot{d}} \times \bold{p} )\times \nabla \times \bold{A})
\end{equation}
To solve the system of equations developed, the weak formulation is used. (See Appendix 2).
The torque developed in the conducting disc is found using (3).
\begin{equation}
T_e = \int_{\Omega_{cc}} [\bold{p} \times (\sigma [(-\nabla \cdot v) + (\bold{\dot{d}}\times \bold{p}) \times (\nabla \times \bold{A})]\times (\nabla \times \bold{A}) )]\cdot \bold{u_z}\ d\Omega
\end{equation}
To address gauge invariance a suitable gauge such as the spanning tree gauge or Coulomb tree gauge can be used\cite{Creuse2019-sz}. This study uses the Coulomb tree guage.
\subsection{Analytic Approach to Design}
The FEM approach of modelling the axial brake makes design difficult since there is no clear link between the equations with the physical and geometrical parameters of the brake. A torque formula has been derived in \cite{Lu2021-lr} for a 3D analytical model under the mean-radius assumption. This formula creates a direct dependency of the torque on the physical and geometrical parameters of the brake. For a full development on the derivation of the Torque see Appendix 3. The Torque formula is given by (4)
\begin{equation}
T_e = \frac{1}{2}\mu_0\ p^2\ \tau R_3\mathfrak{R} \left[\sum_{n=1}^{N}\sum_{k=1}^{K} jk\frac{M^2_{nk}}{a_{nk}} \underline{r}\sinh(a_{nk}b) \right]
\end{equation}
where
\begin{equation}
M_{nk} = \frac{16B_r}{\pi^2\mu_0\ n\ k}\sin(k\alpha\frac{\pi}{2})\sin(n\frac{\pi}{2}\frac{R_2 - R_1}{R_3})
\end{equation}
\begin{equation}\tau = \frac{\pi}{p}R_m\end{equation}
\begin{equation}
R_m = \frac{R_1 + R_2}{2}
\end{equation}
\begin{equation}\underline{r} = \frac{-(\cosh(a_{nk}c)\sinh(\gamma_{nk}d)+\frac{a_{nk}}{\gamma_{nk}}\sinh(a_{nk}c)\cosh(\gamma_{nk}d))}{\cosh(a_{nk}(b+c))\sinh(\gamma_{nk}d)+\frac{a_{nk}}{\gamma_{nk}}\sinh(a_{nk}(b+c))\cosh(\gamma_{nk}d)}\end{equation}
\begin{equation}
a_{nk}=\sqrt{\left(\frac{n\pi}{R_3}\right)^2 + \left(\frac{k\pi}{\tau}\right)^2}
\end{equation}
\begin{equation}\gamma_{nk} = \sqrt{\left(\frac{n\pi}{R_3}\right)^{2}+\left(\frac{k\pi}{\tau}\right)^2 + j\ \sigma\ \mu_0\ N\ R_m\ \frac{k\pi}{\tau}}\end{equation}
The authors of \cite{fontchastagner_2018_design} and \cite{Lu2021-lr} verify that the analytical equations above agree with FEM Simulation results. The link between the FEM analysis and the analytical equations are elaborated in Appendix 3. For a more verbose development of the governing equations, the interested reader is referred to \cite{Lu2021-lr}. When validating these equations, it was difficult to determine exactly what units were used in the development. It seems some are SI and some are not. Hence a scaling factor,$\lambda$, was used to modify (4). The new Torque formula becomes (11)

\begin{equation}
T_e = \frac{\lambda}{2}\mu_0\ p^2\ \tau R_3\mathfrak{R} \left[\sum_{n=1}^{N}\sum_{k=1}^{K} jk\frac{M^2_{nk}}{a_{nk}} \underline{r}\sinh(a_{nk}b) \right]
\end{equation}

\section{Design of Axial Flux Permanent Magnet Eddy Current Braking System}
\subsection{Industry standards}
This work designs adhering around EU/US Standard safety requirements. 

The authors of \cite{li_2011_the} provide braking geometry standards used in the industry and hence the diameter of 280 mm was used for the rotor, is used for the conducting disc considered in this work. The diameter of the magnet mount was be limited to 80\% of the conducting disc giving a diameter of 224 mm similar to the design in \cite{fontchastagner_2018_design}. The thickness constraints of \cite{fontchastagner_2018_design} will be used for the magnet mount. This can be changed later for mechanical stability of the rest of the braking system(mechanical brake, calipers, drivetrain shaft, etc.)

According to \cite{10.2307/44734439}, the United Nations' Economic Commission for Europe(ECE) aspires to harmonize the different European national legislation that currently act as an invisible tariff barrier, and it might affect any European country. Many other European and non-European countries use ECE laws as the basis for their local legislation and approvals. It is reported in \cite{10.2307/44734439} that Sweden's government has pioneered a number of road safety standards, requiring more stringent vehicle/equipment specifications than the ECE, EEC, or any other country. This is especially true in the case of vehicle braking, therefore any assessment of brake restrictions must take the Swedish standards into account.

Per the ECE Regulation 13 \& EEC Directive 71/320. The brake is designed to pass the high-speed test of a modern car which has a maximum speed of 200 $kmh^{-1}$ in most cases. The high-speed test requires the brake to perform at 80\% of $V_{max}$. Furthermore, the Swedish standard\cite{10.2307/44734439}, as well as  UNECE Regulation No 78\cite{economiccommissionforeuropeoftheunitednationsunece_2015_regulation}, requires a minimum acceleration of 4.0 $ms^{-2}$ for lightweight motor vehicles as defined by \cite{10.2307/44734439} which is $\leq$ 3500 kg. 

\begin{equation}
V = 0.8V_{max} = 160\mathrm{kmh^{-1}}
\end{equation}

The friction brake will take over at a low speed such as 20 $kmh^{-1}$\cite{wang_2019_performances}. This means for a deceleration of 4 $\mathrm{ms^-2}$, the brake will take approximately 10s to decelerate to the friction brake threshold. 

The required electromagnetic torque ,$T_e$, given in (13)
\begin{equation}
\begin{split}
T_e &= Fr \\
 & = mar \\
&= 1735*4*0.14 \\
&\approx 972\ \mathrm{N \cdot m}
\end{split}
\end{equation}
Hence the required torque per wheel will be $\approx$ 243 $N\cdot m$.
\subsection{Development of the parameter space}
The design problem becomes searching the parameter space of the brake design shown in \ref{fig9}. subject to constraints $\bold{x}$. Formally we can state this as
\begin{equation}
\begin{split}
T_e(\bold{x}) \geq 243\ \mathrm{N\cdot m}\ \mathrm{for}\ 
1000\mathrm{rpm}\ \leq\ \dot{d}(\bold{x}) \leq 8000\mathrm{rpm} \\ \mathrm{with}: \bold{x} = [R_1,b,w_m]
\end{split}
\end{equation}
The design requirements derived from above are shown in Table 1. 
\begin{table}[!t]
\renewcommand{\arraystretch}{1.3}
\caption{Design Requirements of ECB}
\label{table_example}
\centering
\begin{tabular}{c c c}
\hline
\bfseries Symbol & \bfseries Parameter & \bfseries Value\\
\hline
$T_e$ & Electromagnetic Torque &$243\ \mathrm{N\cdot m}$\\
\hline
$R_3$ & Conducting Disc Radius Torque &$140\ \mathrm{mm}$\\
\hline
$R_2$ & Magnet arrangement outer radius &$112\ \mathrm{mm}$\\
\hline
$w_m$ & Magnet arrangement axial extrusion &$[2,110]\ \mathrm{mm}$\\
\hline
$\dot{d}_{op}$ & Operating Speed &$[1000, 8000]\ \mathrm{rpm}$\\
\hline
$b$ & Magnet Thickness &$[2,40]\ \mathrm{mm}$\\
\hline
\end{tabular}
\end{table}
To simplify the design process, parameters of the brake have been constrained using nominal values from similar work done in \cite{fontchastagner_2018_design} and \cite{oppenheimer_1977_braking}. These constraints are shown in Table 2.

\begin{table}[!t]
\renewcommand{\arraystretch}{1.3}
\caption{Auxiliary Design Constraints of ECB}
\label{table_example}
\centering
\begin{tabular}{c c c}
\hline
\bfseries Symbol & \bfseries Parameter & \bfseries Value\\
\hline
$p$ & Number of pole pairs &4\\
\hline
$\alpha$ & Pole-pitch to pole-arc ratio & 0.444\\
\hline
$c$ & Air-gap thickness &$1\ \mathrm{mm}$\\
\hline
$a,d$ & Back iron thicknesses &$2\ \mathrm{mm}$\\
\hline
$b$ & Magnet Thickness &$[2,40]\ \mathrm{mm}$\\
\hline
\end{tabular}
\end{table}

The material properties chosen for the design were chosen based on what was used for eddy current retarders of similar design goals in \cite{gay_2006_parametric}.

\begin{table}[!t]
\renewcommand{\arraystretch}{1.3}
\caption{Material Properties Used in Design}
\label{table_example}
\centering
\begin{tabular}{c c c}
\hline
\bfseries Symbol & \bfseries Parameter & \bfseries Value\\
\hline
$\mu_0$ & Permeability of free space &$4\pi \times 10^{-7} \mathrm{Hm^{-1}}$\\
\hline
$\sigma$ & Electrical conductivity of disc &$57 \times 10^{7}\ \mathrm{S\dot m}$\\
\hline
$B_r$ & Remenance of PMs &$112\ 1.25\mathrm{T}$\\
\hline
\end{tabular}
\end{table}

\section{Results}
The graphs below show the Torque speed curve for the aforementioned physical and geometrical settings. 
The Torque calculations for the parameters described in Section III are shown for each interval of the RPM range in Fig 4-7. The Torque calculations is plotted over 2kRPM intervals, however, one is done for 2kRPM to investigate if their is any significant variation of Torque at low RPM.

The parameters that closest satisfy the design requirements over all the ranges, along with the parameters that consistently perform close to or the maximum over the ranges are shown in the Table 4.

\begin{table}[!t]
\renewcommand{\arraystretch}{1.3}
\caption{Material Properties Used in Design}
\label{table_example}
\centering
\begin{tabular}{c c c}
\hline
\bfseries Variable & \bfseries Solution 1 & \bfseries Solution 2 \\
\hline
$w_m$(m) & 0.11 & 0.11 \\
\hline
Magnet Thickness, $b$  & 0.03 & 0.04 \\
\hline
$T_e$ @ 1000 rpm & 250.179 & 307.630\\
\hline
$T_e$ @ 2000 rpm& 252.359 & 315.286 \\
\hline
$T_e$ @ 4000 rpm & 253.229 & 315.28 \\
\hline
$T_e$ @ 6000 rpm & 253.229 & 316.177\\
\hline
$T_e$ @ 8000 rpm & 253.802 & 316.710\\
\hline
\end{tabular}
\end{table}

\begin{figure}[!htbp]
\centering
\centering
\includegraphics[width=3.5in]{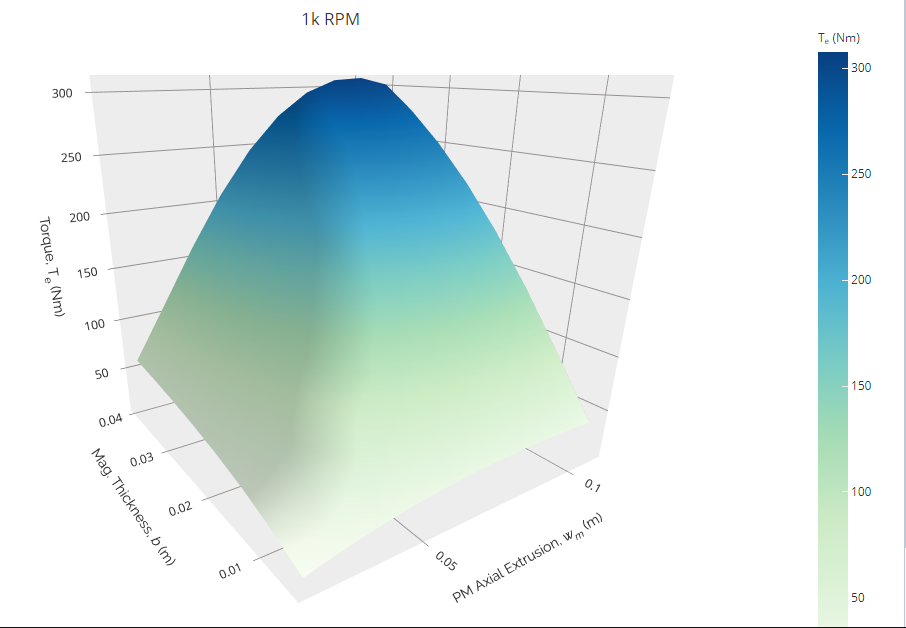}
\caption{Graph showing mesh plot of $T_e,\ b$ and $w_m$ values for 1000 RPM}
\label{fig3}
\end{figure}
\FloatBarrier

\begin{figure}[!htbp]
\centering
\includegraphics[width=3.5in]{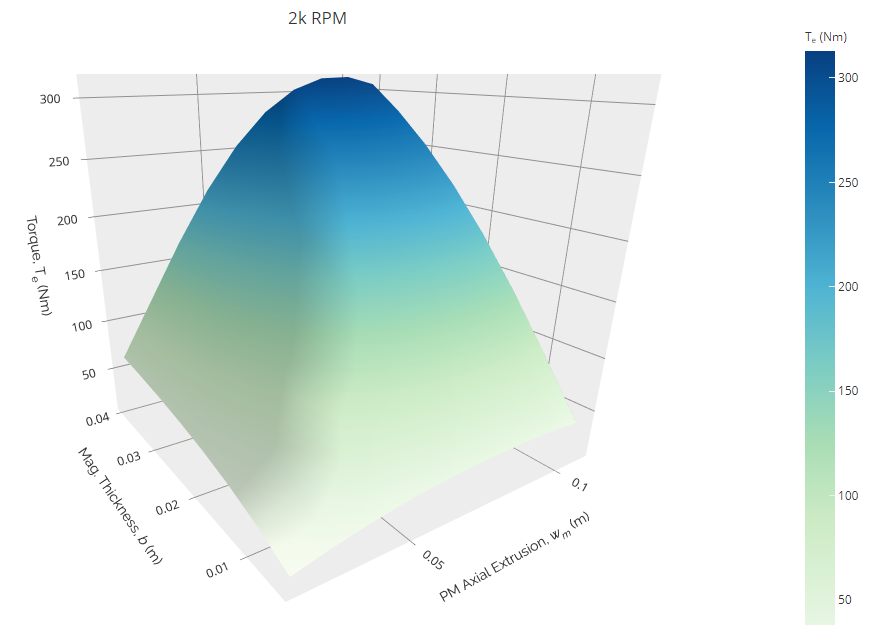}
\caption{Graph showing mesh plot of $T_e,\ b$ and $w_m$ values for 2000 RPM}
\label{fig4}
\end{figure}
\FloatBarrier

\begin{figure}[!htbp]
\centering
\includegraphics[width=3.5in]{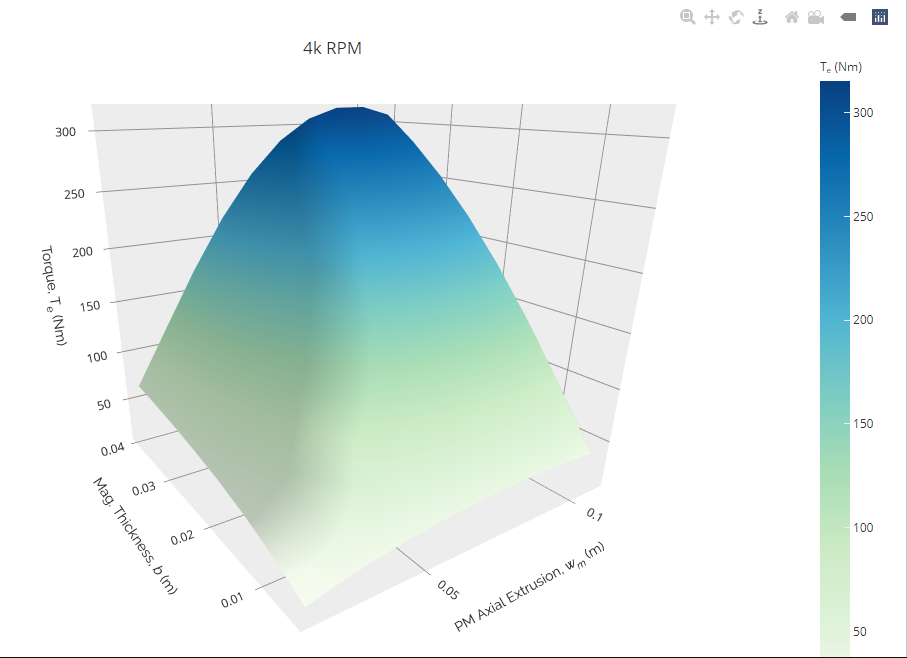}
\caption{Graph showing mesh plot of $T_e,\ b$ and $w_m$ values for 4000 RPM}
\label{fig5}
\end{figure}
\FloatBarrier

\begin{figure}[!htbp]
\centering
\includegraphics[width=3.5in]{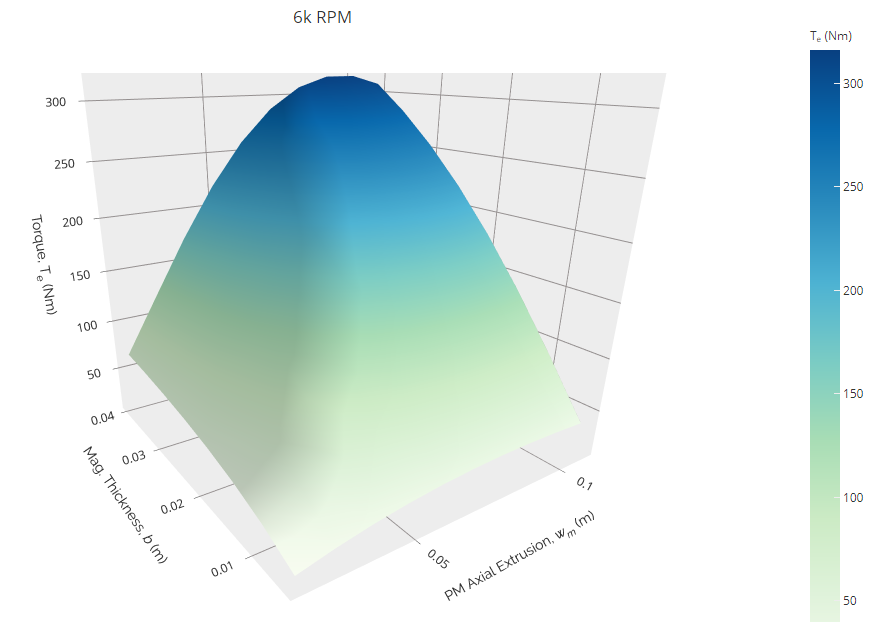}
\caption{Graph showing mesh plot of $T_e,\ b$ and $w_m$ values for 6000 RPM}
\label{fig6}
\end{figure}
\FloatBarrier

\begin{figure}[!htbp]
\centering
\includegraphics[width=3.5in]{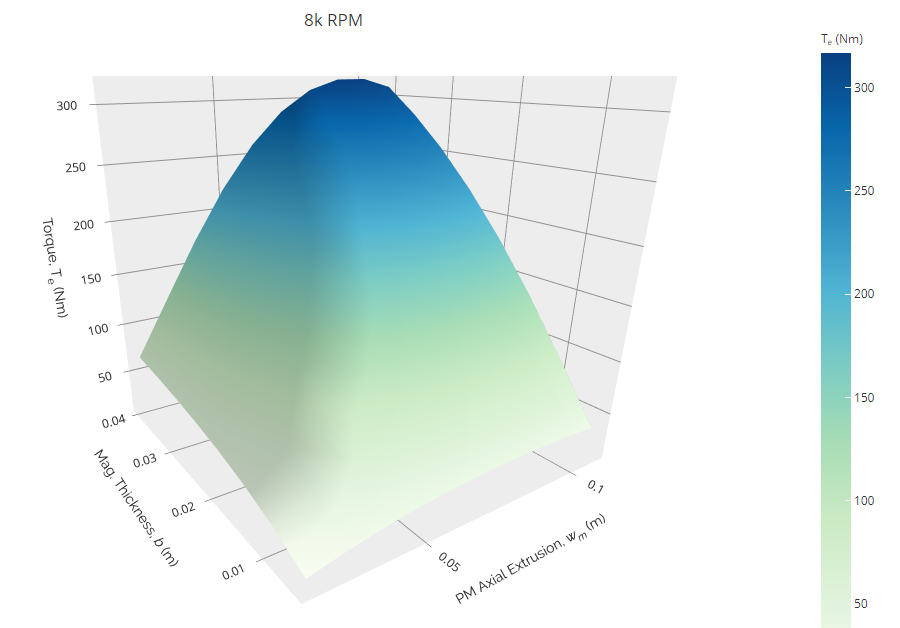}
\caption{Graph showing mesh plot of $T_e,\ b$ and $w_m$ values for 8000 RPM}
\label{fig7}
\end{figure}
\FloatBarrier

\section{Discussion}
The modified Torque design equations have been verified by replicating the results of \cite{fontchastagner_2018_design} and \cite{gay_2006_parametric}. Thus, this work reveals a less ambiguous application of the analytical approach done in \cite{Lu2021-lr} as all calculations are done in SI units and explicitly stated in the open source code provided by the author. To meet size constraints, the Torque was distributed among for wheels and the design requirement was constructed. 

From the graphs above, the Torque does not significantly vary over the RPM for $b-w_m$ parameter combinations. It is however seen from Table IV that the magnet thickness has a larger influence on the Torque. Thus if the design needs to be extended for greater Torque applications, a thicker surface mount magnet can be used. The closest Torque achieved to the target was 253.802 rpm with a magnet thickness of 0.03m and and 0.11cm radial extrusion.

The findings in this work agree closely with the calculated Torque values for a similar geometry in \cite{gay_2006_parametric} which used FEM simulations to verify their findings. Hence it can be said that the values should perform well in FEM validation

\section{Conclusion}
This study utilizes a distributed permanent magnet axial flux braking system to satisfy industry standards for braking of a light weight motor vehicle. Analytical equations guided the design and common part sizes were used guide the parameter search. It was found that a magnet thickness of 0.03 RPM and 0.11cm radial extrusion allowed for the closest match to the UNECE braking standards. This study is limited to the accuracy of the analytic equations used in \cite{Lu2021-lr} and  future work can be done to account for what parameters contribute to the scaling factor $\lambda$ needed to meet the design deliverable. All data and findings can be found via the GitHub repository used for the work\footnote{https://github.com/shanks847/Axial-Flux-Permanent-Magnet-Eddy-Current-Brake}

%
%
\bibliographystyle{styles/spmpsci_unsrt}
\bibliography{references}

\newpage
\section*{Appendix I}
This Appendix contains bulky images that could not be included in the body of the report.

\begin{figure}[!htp]
\centering
\includegraphics[width=3.5in]{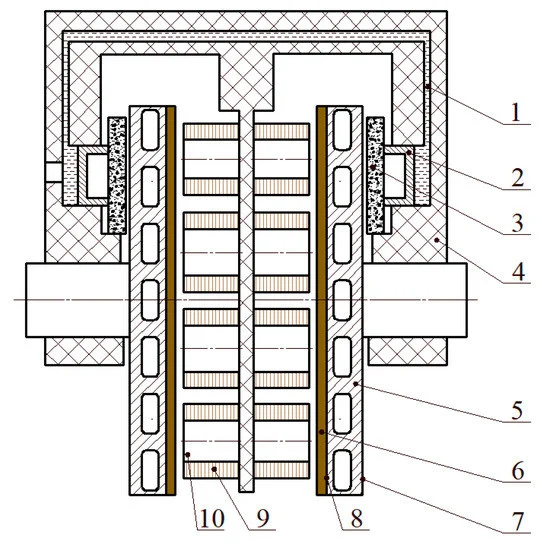}
\caption{Structural diagram of the electromagnetic-frictional integrated brake. 1, brake fluid; 2, brake piston; 3, brake pad; 4, caliper body; 5, integrated brake disc; 6, copper layer; 7, friction brake surface; 8, electromagnetic brake surface; 9, coil; and 10, iron core, Taken from \cite{wang_2019_performances}}
\label{fig8}
\end{figure}

\begin{figure}[!htp]
\centering
\includegraphics[width=3.5in]{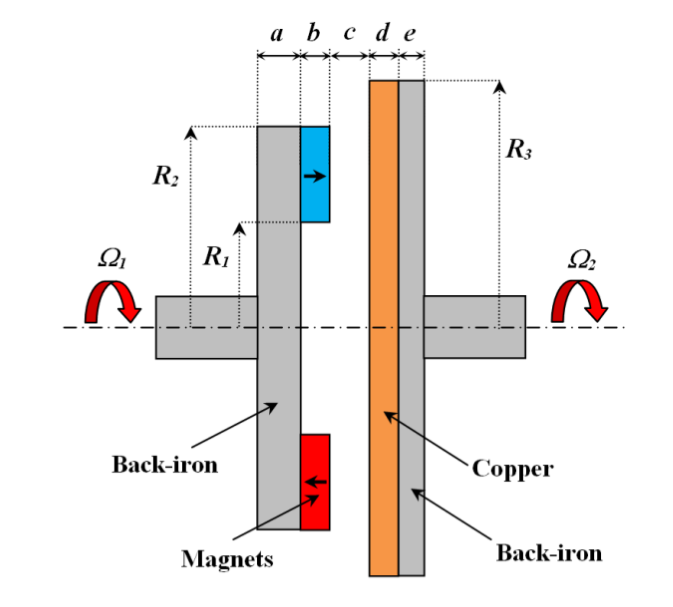}
\caption{Diagram of braking system showing geometrical parameters. (Modified from \cite{fontchastagner_2018_design})}
\label{fig9}
\end{figure}
\FloatBarrier

\section*{Appendix II}
A development of the FEM approach to the design problem is shown here. This is a modified version of the work done in \cite{fontchastagner_2017_axialfield}. It is included to make the link between the FEM approach and Analytical clearer. The principles remain the same but different notation and simplifications are used to maintain homogeneity.

In steady-state, the driving rotor rotates at speed $\dot{d}_1$ and the driven one at speed $\dot{d}_2$. Choosing the reference frame $(\mathrm{O},u_x,u_y,u_z)$ link to the first rotor, the second rotor rotates at speed $\dot{d}  = \dot{d}_2 - \dot{d}_1$ which is the slip speed of the coupling. All other components are stationary. The $z$ axis is defined as the axis of rotation and thus the rotation vector is $\mathbf{\dot{d}} = \dot{d}u_z$. This work uses the same geometry configuration as \cite{fontchastagner_2017_axialfield} and hence the domains and boundaries of the geometry are labelled similar to original authors.

The magnet remanence are defined by (15)
\begin{equation}
\bold{B_r} = \pm B_r\bold{u_z}
\end{equation}
Let the position vector be $\bold{p}$
\begin{equation}
\bold{p} = (x,y,z)
\end{equation}
The speed, $\bold{v}$, at the position, $\bold{p}$ is given in (17)

\begin{equation}
\bold{v} = \bold{\dot{d}}\times \bold{p}
\end{equation}

The partial derivative of the flux density is given as

\begin{equation}
\frac{\partial \bold{B}}{\partial t} = -\nabla \times (\bold{v} \times \bold{B})
\end{equation}

Thus, the Maxwell-Faraday equation becomes (19)

\begin{equation}
\nabla \times (\bold{E} + \bold{v} \times \bold{B}) = \nabla \times (\bold{E} + (\bold{\dot{d}} \times \bold{p}) \times \bold{B} ) = 0
\end{equation}

The problem can be solved via quasi-static solutions from classical magnetostatics. The two possible solution methods are using the magnetic vector potential or the magnetic scale potential. According to the authors of the original work, the magnetic vector potential provides more accurate modeling hence it will be used in this work.

By using the modified form of Ohms Law with the magnetic vector potential, and the the scalar electric potential $v$, the eddy currents in the conducting elements are given by (19)

\begin{equation}
\bold{J_e} = \sigma(\ (-\nabla\cdot \phi) + (\bold{\dot{d}} \times \bold{p} )\times \nabla \times \bold{A})
\end{equation}

The total domain, $\Omega$, the whole conducting domain $\Omega_{cc}$ and its boundaries $\Gamma_c$ are defined by (21)

\begin{equation}
\begin{aligned}
\Omega\ \ &= \Omega_m \cup \Omega_{ym} \cup \Omega_c \cup \Omega_{yc} \cup \Omega_a\\ \ \Omega_{cc} &= \Omega_c \cup \Omega_{yc}\\ \Gamma_c\  &= \partial \Omega_{cc}
\end{aligned}
\end{equation}

The complete variational formulation of the $A-\phi$ formulation is given by (22)

\begin{equation}
\begin{aligned} 
\begin{cases}
(\mu^{-1}\nabla \times \bold{A},\nabla \times \bold{A'})_{\Omega}-(\mu^{-1}\bold{B_r},\nabla \times \bold{A'})_{\Omega_m} +\\ (\sigma \nabla \phi,\bold{A'})_{\Omega_{cc}} - (\sigma(\bold{\dot{d}}\times \bold{p}) \times \nabla \times \bold{A}, \bold{A'}) = 0\\
(\nabla \cdot \phi,\nabla \cdot \phi')_{\Omega_{cc}} - (\sigma(\bold{\dot{d}}\times \bold{p} \times (\nabla \times \bold{A}), \nabla \cdot \phi')_{\Omega_{cc}}) = 0 \\
\forall \bold{A'} \in \bold{H_0} (\nabla \times,\Omega), \forall \phi' \in H_0^1(\Omega_{cc})
\end{cases} 
\end{aligned}
\end{equation} 

The eddy current density in the domain $\Omega_{cc}$ is obtained from (20). The power dissipation, $P_J$, can be computed by (23)

\begin{equation}
P_J = \int_{\Omega_{cc}} \sigma || (\nabla\cdot \phi -(\bold{\dot{d}} \times p) \times (\nabla \times A)||^2\  d\Omega
\end{equation}

The electromagnetic torque, $T_{e}$ within $\Omega_{cc}$ can be computed using Laplace forces utilizing the magnetic vector potential and scalar electric potential.
\begin{equation}
T_{e} = \int_{\Omega_{cc}} [\bold{p} \times (\sigma [(-\nabla \cdot \phi) + (\bold{\dot{d}}\times \bold{p}) \times (\nabla \times \bold{A})]\times (\nabla \times \bold{A}) )]\cdot \bold{u_z}\ d\Omega
\end{equation}

To ensure the uniqueness of $\bold{A}$, the gauge invariance must be addressed. A spanning tree gage \cite{fontchastagner_2017_axialfield},\cite{dular_1995_a} is used and a Dirichlet condition \cite{Creuse2019-sz} upon $\bold{A}$ is applied on $\Gamma_d$.

\section*{Appendix III}
The FEM equations developed in Appendix 2 are developed further to link them to the Analytic equations used in the design.

The $A-\phi$ variational formulation explains how to solve the equations developed above and is found in \cite{fontchastagner_2017_axialfield}. Supporting material on weak formulations, function spaces, and Maxwell’s Time-Harmonic form used in \cite{fontchastagner_2017_axialfield} is found in \cite{kagerer_2018_finite} and \cite{kanayama_2000_an}. 
The segment notation used in the  single pole sector used in \cite{Lu2021-lr} are used here. 
To chain together the analysis from \cite{fontchastagner_2017_axialfield} to \cite{Lu2021-lr}, the regions of the geometry are denoted with slightly different subscripts. Let Region I be $\Omega_{ym}$, Region II be $\Omega_m$, Region III be $\Omega_a$, Region IV be $\Omega_c$ and Region V be $\Omega_{yc}$. The flux density and magnetic field intensity in regions, I, II \& III  are given by:

\begin{equation}
\begin{aligned} 
\begin{cases}
\bold{B}_I &= \mu_0\mu_{iron}\bold{H}_1\\
\bold{B}_{II} &= \mu_0\mu_r\bold{H}_{11} + \mu_0\bold{M}\\
\bold{B}_{III} &= \mu_0\bold{H}_{III}
\end{cases} 
\end{aligned}
\end{equation}
In region IV, the time-invariant steady state is considered. The magnetic field satisfies the static approximation of Maxwell’s equations as:
\begin{equation}
\begin{aligned} 
\begin{cases}
\nabla\cdot  \bold{B}_{IV} &= 0\\
\nabla \times \bold{H}_{IV}&= \bold{J}_{IV}\\
\nabla \times \bold{E}_{IV} &= \frac{-\partial \bold{B}_{IV}}{\partial t} = 0
\end{cases} 
\end{aligned}
\end{equation}
The current density in the moving plate is expressed as
\begin{equation}
J_{IV} = \sigma(\bold{E}_{IV}+v_m\times \bold{B}_{IV})
\end{equation}

Equations (26) and (27) are solved using an $A-\phi$ variational formulation.

Using Kirchoff’s Law the governing equations for each domain are found in \cite{Lu2021-lr}, \cite{fontchastagner_2018_design}, and \cite{nehl_1994_nonlinear}.

The governing equations and boundary conditions in the finite element analysis are:

\begin{equation}
\begin{aligned} 
\begin{cases}
\nabla \times \mu_r^{-1}\nabla\times \bold{A}-\nabla\mu_{iron}^{-1}\nabla\cdot bold{A}&= \begin{aligned}
\begin{cases}
\bold{J}_{IV}\ \mathrm{in\ region\ IV}\\0\ \mathrm{in\ region\ III}\\\nabla\times M\ \mathrm{in\ region\ II}
\end{cases}
\end{aligned}\\
S_{boundary}
\begin{aligned}
\begin{cases}
\bold{A}\times \bold{n} &= 0\\
\mu_{iron}^{-1}\nabla\cdot \bold{A} &= 0
\end{cases} 
\end{aligned}
\end{cases}
\end{aligned}
\end{equation}

The electromagnetic torque, $T_e$, is calculated as:

\begin{equation}
\begin{aligned}
T_e &= \bold{F}\cdot r\\
&= r\int_{Disc}\bold{J} \times \bold{B}\ dV
\end{aligned}
\end{equation}

An H-formulation is used to solve the eddy current problem in region IV under the magneto-static case as 
\begin{equation}
 \nabla^{2}\bold{H}_{IV}=-\sigma\mu_0\nabla\times(v_m \times \bold{H}_{IV})   
\end{equation}

The method of separation of variables to solve for the magnetic scalar potential and magnetic intensity in the xyz plane for the regions I-IV\cite{Lu2021-lr}.
Recalling

\begin{equation}
\bold{H} = \mu_0\bold{B}
\end{equation}

Then by integrating the magnetic field over the Disc, the electromagnetic torque can be calculated as shown in \cite{Lu2021-lr}. Expanding the electromagnetic torque equation in \cite{Lu2021-lr} gives (4). It is worth noting that the authors of \cite{fontchastagner_2018_design} use an $A-\phi$ formulation to solve the magneto-static equations used to develop the electromagnetic torque formula and found there was little loss of accuracy hence the analytical expressions can be used to design systems accurately and then improved with FEM.
\end{document}